\documentclass{svmult}
\usepackage{epsfig,graphicx} 
\usepackage[bottom]{footmisc}
\usepackage{natbib,url}      
\usepackage{amssymb}      
\bibpunct{(}{)}{;}{a}{}{,}   
\def\cite#1{\citealp{#1}}    
\def\authorindex#1{}  

\begin{document}\newcount\preprintheader\preprintheader=1


\def\thisvolume{these proceedings}

\def\aj{{AJ}}			
\def\araa{{ARA\&A}}		
\def\apj{{ApJ}}			
\def\apjl{{ApJ}}		
\def\apjs{{ApJS}}		
\def\ao{{Appl.\ Optics}} 
\def\apss{{Ap\&SS}}		
\def\aap{{A\&A}}		
\def\aapr{{A\&A~Rev.}}		
\def\aaps{{A\&AS}}		
\def\an{{Astron.\ Nachrichten}}
\def\aspcs{{ASP Conf.\ Ser.}}
\def\assp{{Astrophys.\ \& Space Sci.\ Procs., Springer, Heidelberg}}
\def\azh{{AZh}}			
\def\baas{{BAAS}}		
\def\jrasc{{JRASC}}	
\def\memras{{MmRAS}}		
\def\mnras{{MNRAS}}
\def\nat{{Nat}}		
\def\pra{{Phys.\ Rev.\ A}} 
\def\prb{{Phys.\ Rev.\ B}}		
\def\prc{{Phys.\ Rev.\ C}}		
\def\prd{{Phys.\ Rev.\ D}}		
\def\prl{{Phys.\ Rev.\ Lett.}} 
\def\pasp{{PASP}}
\def\pasj{{PASJ}}		
\def\qjras{{QJRAS}}
\def\science{{Sci}}		
\def\skytel{{S\&T}}		
\def\solphys{{Solar\ Phys.}} 
\def\sovast{{Soviet\ Ast.}}  
\def\ssr{{Space\ Sci.\ Rev.}}
\def\svassp{{Astrophys.\ Space Sci.\ Procs., Springer, Heidelberg}}
\def\zap{{ZAp}}			
\let\astap=\aap
\let\apjlett=\apjl
\let\apjsupp=\apjs
\def\grl{{Geophys.\ Res.\ Lett.}}  
\def\jgr{{J. Geophys.\ Res.}} 

\def\ion#1#2{{\rm #1}\,{\uppercase{#2}}}  
\def\deg{\hbox{$^\circ$}}
\def\sun{\hbox{$\odot$}}
\def\earth{\hbox{$\oplus$}}
\def\la{\mathrel{\hbox{\rlap{\hbox{\lower4pt\hbox{$\sim$}}}\hbox{$<$}}}}
\def\ga{\mathrel{\hbox{\rlap{\hbox{\lower4pt\hbox{$\sim$}}}\hbox{$>$}}}}
\def\sq{\hbox{\rlap{$\sqcap$}$\sqcup$}}
\def\arcmin{\hbox{$^\prime$}}
\def\arcsec{\hbox{$^{\prime\prime}$}}
\def\fd{\hbox{$.\!\!^{\rm d}$}}
\def\fh{\hbox{$.\!\!^{\rm h}$}}
\def\fm{\hbox{$.\!\!^{\rm m}$}}
\def\fs{\hbox{$.\!\!^{\rm s}$}}
\def\fdg{\hbox{$.\!\!^\circ$}}
\def\farcm{\hbox{$.\mkern-4mu^\prime$}}
\def\farcs{\hbox{$.\!\!^{\prime\prime}$}}
\def\fp{\hbox{$.\!\!^{\scriptscriptstyle\rm p}$}}
\def\micron{\hbox{$\mu$m}}
\def\onehalf{\hbox{$\,^1\!/_2$}}	
\def\onethird{\hbox{$\,^1\!/_3$}}
\def\twothirds{\hbox{$\,^2\!/_3$}}
\def\onequarter{\hbox{$\,^1\!/_4$}}
\def\threequarters{\hbox{$\,^3\!/_4$}}
\def\ubv{\hbox{$U\!BV$}}		
\def\ubvr{\hbox{$U\!BV\!R$}}		
\def\ubvri{\hbox{$U\!BV\!RI$}}		
\def\ubvrij{\hbox{$U\!BV\!RI\!J$}}		
\def\ubvrijh{\hbox{$U\!BV\!RI\!J\!H$}}		
\def\ubvrijhk{\hbox{$U\!BV\!RI\!J\!H\!K$}}		
\def\ub{\hbox{$U\!-\!B$}}		
\def\bv{\hbox{$B\!-\!V$}}		
\def\vr{\hbox{$V\!-\!R$}}		
\def\ur{\hbox{$U\!-\!R$}}


\def\labelitemi{{\bf --}}  

\def\rmit#1{{\it #1}}              
\def\rmit#1{{\rm #1}}              
\def\etal{\rmit{et al.}}           
\def\etc{\rmit{etc.}}           
\def\ie{\rmit{i.e.,}}              
\def\eg{\rmit{e.g.,}}              
\def\cf{cf.}                       
\def\viz{\rmit{viz.}}
\def\vs{\rmit{vs.}}

\def\rot{\hbox{\rm rot}}
\def\div{\hbox{\rm div}}
\def\lesssim{\mathrel{\hbox{\rlap{\hbox{\lower4pt\hbox{$\sim$}}}\hbox{$<$}}}}
\def\gtrsim{\mathrel{\hbox{\rlap{\hbox{\lower4pt\hbox{$\sim$}}}\hbox{$>$}}}}
\def\dif{\: {\rm d}}                       
\def\ep{\:{\rm e}^}                        
\def\dash{\hbox{$\,-\,$}}                  
\def\is{\!=\!}                             

\def\starname#1#2{${#1}$\,{\rm {#2}}}  
\def\Teff{\hbox{$T_{\rm eff}$}}   

\def\kms{\hbox{km$\;$s$^{-1}$}}
\def\ms{\hbox{m$\;$s$^{-1}$}}
\def\Mxcm{\hbox{Mx\,cm$^{-2}$}}    

\def\Bapp{\hbox{$B_{\rm app}$}}    

\def\komega{($k, \omega$)}                 
\def\kf{($k_h,f$)}                         
\def\VminI{\hbox{$V\!\!-\!\!I$}}           
\def\IminI{\hbox{$I\!\!-\!\!I$}}           
\def\VminV{\hbox{$V\!\!-\!\!V$}}           
\def\Xt{\hbox{$X\!\!-\!t$}}                

\def\level #1 #2#3#4{$#1 \: ^{#2} \mbox{#3} ^{#4}$}   

\def\specchar#1{\uppercase{#1}}    
\def\AlI{\mbox{Al\,\specchar{i}}}  
\def\BI{\mbox{B\,\specchar{i}}} 
\def\BII{\mbox{B\,\specchar{ii}}}  
\def\BaI{\mbox{Ba\,\specchar{i}}}  
\def\BaII{\mbox{Ba\,\specchar{ii}}} 
\def\CI{\mbox{C\,\specchar{i}}} 
\def\CII{\mbox{C\,\specchar{ii}}} 
\def\CIII{\mbox{C\,\specchar{iii}}} 
\def\CIV{\mbox{C\,\specchar{iv}}} 
\def\CaI{\mbox{Ca\,\specchar{i}}} 
\def\CaII{\mbox{Ca\,\specchar{ii}}} 
\def\CaIII{\mbox{Ca\,\specchar{iii}}} 
\def\CoI{\mbox{Co\,\specchar{i}}} 
\def\CrI{\mbox{Cr\,\specchar{i}}} 
\def\CriI{\mbox{Cr\,\specchar{ii}}} 
\def\CsI{\mbox{Cs\,\specchar{i}}} 
\def\CsII{\mbox{Cs\,\specchar{ii}}} 
\def\CuI{\mbox{Cu\,\specchar{i}}} 
\def\FeI{\mbox{Fe\,\specchar{i}}} 
\def\FeII{\mbox{Fe\,\specchar{ii}}} 
\def\FeIX{\mbox{Fe\,\specchar{ix}}}
\def\FeX{\mbox{Fe\,\specchar{x}}}
\def\FeXVI{\mbox{Fe\,\specchar{xvi}}}
\def\FrI{\mbox{Fr\,\specchar{i}}}
\def\HI{\mbox{H\,\specchar{i}}} 
\def\HII{\mbox{H\,\specchar{ii}}} 
\def\Hmin{\hbox{\rmH$^{^{_{\scriptstyle -}}}$}}      
\def\Hemin{\hbox{{\rm He}$^{^{_{\scriptstyle -}}}$}} 
\def\HeI{\mbox{He\,\specchar{i}}} 
\def\HeII{\mbox{He\,\specchar{ii}}} 
\def\HeIII{\mbox{He\,\specchar{iii}}} 
\def\KI{\mbox{K\,\specchar{i}}} 
\def\KII{\mbox{K\,\specchar{ii}}} 
\def\KIII{\mbox{K\,\specchar{iii}}} 
\def\LiI{\mbox{Li\,\specchar{i}}} 
\def\LiII{\mbox{Li\,\specchar{ii}}} 
\def\LiIII{\mbox{Li\,\specchar{iii}}} 
\def\MgI{\mbox{Mg\,\specchar{i}}} 
\def\MgII{\mbox{Mg\,\specchar{ii}}} 
\def\MgIII{\mbox{Mg\,\specchar{iii}}} 
\def\MnI{\mbox{Mn\,\specchar{i}}} 
\def\NI{\mbox{N\,\specchar{i}}}
\def\NIV{\mbox{N\,\specchar{iv}}}
\def\NaI{\mbox{Na\,\specchar{i}}}
\def\NaII{\mbox{Na\,\specchar{ii}}}
\def\NaIII{\mbox{Na\,\specchar{iii}}}
\def\NeVIII{\mbox{Ne\,\specchar{viii}}} 
\def\NiI{\mbox{Ni\,\specchar{i}}} 
\def\NiII{\mbox{Ni\,\specchar{ii}}}
\def\NiIII{\mbox{Ni\,\specchar{iii}}} 
\def\OI{\mbox{O\,\specchar{i}}} 
\def\OVI{\mbox{O\,\specchar{vi}}}
\def\RbI{\mbox{Rb\,\specchar{i}}} 
\def\SII{\mbox{S\,\specchar{ii}}} 
\def\SiI{\mbox{Si\,\specchar{i}}} 
\def\SiII{\mbox{Si\,\specchar{ii}}} 
\def\SrI{\mbox{Sr\,\specchar{i}}}
\def\SrII{\mbox{Sr\,\specchar{ii}}}
\def\TiI{\mbox{Ti\,\specchar{i}}} 
\def\TiII{\mbox{Ti\,\specchar{ii}}} 
\def\TiIII{\mbox{Ti\,\specchar{iii}}} 
\def\TiIV{\mbox{Ti\,\specchar{iv}}} 
\def\VI{\mbox{V\,\specchar{i}}} 
\def\HtwoO{\mbox{H$_2$O}}        
\def\Otwo{\mbox{O$_2$}}          

\def\Halpha{\mbox{H\hspace{0.1ex}$\alpha$}} 
\def\Ha{\mbox{H\hspace{0.2ex}$\alpha$}}
\def\Hbeta{\mbox{H\hspace{0.2ex}$\beta$}}
\def\Hgamma{\mbox{H\hspace{0.2ex}$\gamma$}}
\def\Hdelta{\mbox{H\hspace{0.2ex}$\delta$}}
\def\Hepsilon{\mbox{H\hspace{0.2ex}$\epsilon$}}
\def\Hzeta{\mbox{H\hspace{0.2ex}$\zeta$}}
\def\Lyalpha{\mbox{Ly$\hspace{0.2ex}\alpha$}}
\def\Lybeta{\mbox{Ly$\hspace{0.2ex}\beta$}}
\def\Lygamma{\mbox{Ly$\hspace{0.2ex}\gamma$}}
\def\Lycont{\mbox{Ly\hspace{0.2ex}{\small cont}}}
\def\Baalpha{\mbox{Ba$\hspace{0.2ex}\alpha$}}
\def\Babeta{\mbox{Ba$\hspace{0.2ex}\beta$}}
\def\Bacont{\mbox{Ba\hspace{0.2ex}{\small cont}}}
\def\Paalpha{\mbox{Pa$\hspace{0.2ex}\alpha$}}
\def\Bralpha{\mbox{Br$\hspace{0.2ex}\alpha$}}

\def\NaD{\mbox{Na\,\specchar{i}\,D}}    
\def\NaDone{\mbox{Na\,\specchar{i}\,\,D$_1$}}
\def\NaDtwo{\mbox{Na\,\specchar{i}\,\,D$_2$}}
\def\NaID{\mbox{Na\,\specchar{i}\,\,D}}
\def\NaIDone{\mbox{Na\,\specchar{i}\,\,D$_1$}}
\def\NaIDtwo{\mbox{Na\,\specchar{i}\,\,D$_2$}}
\def\Done{\mbox{D$_1$}}
\def\Dtwo{\mbox{D$_2$}}

\def\Mgbone{\mbox{Mg\,\specchar{i}\,b$_1$}}
\def\Mgbtwo{\mbox{Mg\,\specchar{i}\,b$_2$}}
\def\Mgbthree{\mbox{Mg\,\specchar{i}\,b$_3$}}
\def\MgIb{\mbox{Mg\,\specchar{i}\,b}}
\def\MgIbone{\mbox{Mg\,\specchar{i}\,b$_1$}}
\def\MgIbtwo{\mbox{Mg\,\specchar{i}\,b$_2$}}
\def\MgIbthree{\mbox{Mg\,\specchar{i}\,b$_3$}}

\def\CaIIK{\mbox{Ca\,\specchar{ii}\,K}}       
\def\CaIIH{\mbox{Ca\,\specchar{ii}\,H}}
\def\CaIIHK{\mbox{Ca\,\specchar{ii}\,H\,\&\,K}}
\def\HK{\mbox{H\,\&\,K}}
\def\Kthree{\mbox{K$_3$}}      
\def\Hthree{\mbox{H$_3$}}
\def\Ktwo{\mbox{K$_2$}}
\def\Htwo{\mbox{H$_2$}}
\def\Kone{\mbox{K$_1$}}     
\def\Hone{\mbox{H$_1$}}     
\def\KtwoV{\mbox{K$_{2V}$}}
\def\KtwoR{\mbox{K$_{2R}$}}
\def\KoneV{\mbox{K$_{1V}$}}
\def\KoneR{\mbox{K$_{1R}$}}
\def\HtwoV{\mbox{H$_{2V}$}}
\def\HtwoR{\mbox{H$_{2R}$}}
\def\HoneV{\mbox{H$_{1V}$}}
\def\HoneR{\mbox{H$_{1R}$}}

\def\hk{\mbox{h\,\&\,k}}
\def\kthree{\mbox{k$_3$}}    
\def\hthree{\mbox{h$_3$}}
\def\ktwo{\mbox{k$_2$}}
\def\htwo{\mbox{h$_2$}}
\def\kone{\mbox{k$_1$}}     
\def\hone{\mbox{h$_1$}}     
\def\ktwoV{\mbox{k$_{2V}$}}
\def\ktwoR{\mbox{k$_{2R}$}}
\def\koneV{\mbox{k$_{1V}$}}
\def\koneR{\mbox{k$_{1R}$}}
\def\htwoV{\mbox{h$_{2V}$}}
\def\htwoR{\mbox{h$_{2R}$}}
\def\honeV{\mbox{h$_{1V}$}}
\def\honeR{\mbox{h$_{1R}$}}

\ifnum\preprintheader=1     
\makeatletter  
\def\@maketitle{\newpage
\markboth{}{}%
  {\mbox{} \vspace*{-8ex} \par 
   \em \footnotesize To appear in ``Magnetic Coupling between the Interior 
       and the Atmosphere of the Sun'', eds. S.~S.~Hasan and R.~J.~Rutten, 
       Astrophysics and Space Science Proceedings, Springer-Verlag, 
       Heidelberg, Berlin, 2009.} \vspace*{-5ex} \par
 \def\lastand{\ifnum\value{@inst}=2\relax
                 \unskip{} \andname\
              \else
                 \unskip \lastandname\
              \fi}%
 \def\and{\stepcounter{@auth}\relax
          \ifnum\value{@auth}=\value{@inst}%
             \lastand
          \else
             \unskip,
          \fi}%
  \raggedright
 {\Large \bfseries\boldmath
  \pretolerance=10000
  \let\\=\newline
  \raggedright
  \hyphenpenalty \@M
  \interlinepenalty \@M
  \if@numart
     \chap@hangfrom{}
  \else
     \chap@hangfrom{\thechapter\thechapterend\hskip\betweenumberspace}
  \fi
  \ignorespaces
  \@title \par}\vskip .8cm
\if!\@subtitle!\else {\large \bfseries\boldmath
  \vskip -.65cm
  \pretolerance=10000
  \@subtitle \par}\vskip .8cm\fi
 \setbox0=\vbox{\setcounter{@auth}{1}\def\and{\stepcounter{@auth}}%
 \def\thanks##1{}\@author}%
 \global\value{@inst}=\value{@auth}%
 \global\value{auco}=\value{@auth}%
 \setcounter{@auth}{1}%
{\lineskip .5em
\noindent\ignorespaces
\@author\vskip.35cm}
 {\small\institutename\par}
 \ifdim\pagetotal>157\p@
     \vskip 11\p@
 \else
     \@tempdima=168\p@\advance\@tempdima by-\pagetotal
     \vskip\@tempdima
 \fi
}
\makeatother     
\fi

\title*{Three-Dimensional Magnetic Reconnection}


\author{C. E. Parnell \and A. L. Haynes}

\authorindex{Parnell, C. E.} 
\authorindex{Haynes, A. L.} 

\institute{School of Mathematics \& Statistics, University of St Andrews, Scotland}

\maketitle

\setcounter{footnote}{0}  

\begin{abstract} 
  The importance of magnetic reconnection as an energy release mechanism in many solar, stellar, magnetospheric and astrophysical phenomena has long been recognised. Reconnection is the only mechanism by which magnetic fields can globally restructure, enabling them to access a lower energy state. Over the past decade, there have been some major advances in our understanding of three-dimensional reconnection. In particular, the key characteristics of 3D magnetohydrodynamic (MHD) reconnection have been determined. For instance, 3D reconnection (i) occurs with or without nulls, (ii) occurs continuously and continually throughout a diffusion region and (iii) is driven by counter rotating flows. 

Furthermore, analysis of resistive 3D MHD magnetic experiments have revealed some intriguing effects relating to where and how reconnection occurs. To illustrate these new features, a series of constant-resistivity experiments, involving the interaction of two opposite-polarity magnetic sources in an overlying field, are considered. Such a simple interaction represents a typical building block of the Sun's magnetic atmosphere. By following the evolution of the magnetic topology, we are able to explain where, how and at what rate the reconnection occurs. Remarkably there can be up to five energy release sites at anyone time (compared to one in the potential case) and the duration of the interaction increases (more than doubles) as the resistivity decreases (by a factor of 16). The decreased resistivity also leads to a higher peak ohmic dissipation and more energy being released in total, as a result of a greater injection of Poynting flux.
\end{abstract}

\section{Introduction}      \label{parnell-sec:introduction}

Magnetic reconnection is a fundamental plasma physics process that is essential to many phenomena on the Sun, such as solar flares, CMEs, coronal heating, nano/microflares, X-ray bright points, explosive events and the solar dynamo. It is also an extremely important mechanism in the magnetosphere where it plays a key role in linking the magnetic fields from the Sun and Earth and in powering flux transfer events and substorms. It is also very important in many astrophysics applications such as accretion discs, stellar flares and coron\ae, astrophysical jets and stellar dynamos.

Magnetic reconnection is important for two key reasons: First, it is a mechanism by which energy stored in a magnetic field may be rapidly released and converted into thermal and kinetic energy, causing heating, bulk plasma motions and the acceleration of particles. Secondly, it is the mechanism by which global restructuring of the magnetic field may take place. Indeed, it is this restructuring that facilitates the release of energy by allowing the magnetic field to access a lower energy state. 

\begin{figure}
\centering{
  \includegraphics[width=4.9cm]{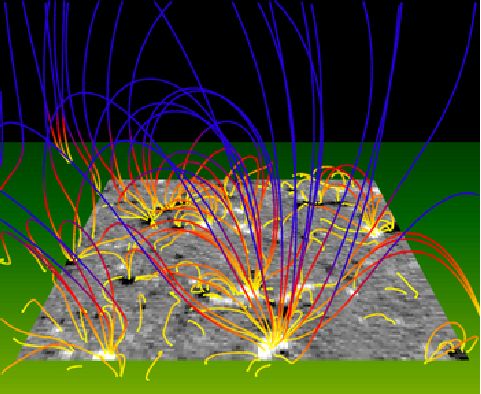}}
\caption[]{Potential magnetic field extrapolation from a quiet-Sun magnetogram. \citep{cep-Close2004}.
\label{qsmagfield}}
\end{figure}

The entire surface of the Sun is threaded by magnetic fields that are directed both into and out of the Sun. The magnetic field is clumped into numerous photospheric flux features that range from large sunspots with fluxes of about $10^{20}$ Mx \citep{cep-SchrijverH1994} down to tiny intranetwork fields with just $10^{16}$ Mx or less \citep{cep-Wang1995}. These photospheric features, both small and large, are the feet of magnetic loops that are intermingled and expand to fill the whole of the solar atmosphere (Fig. ~\ref{qsmagfield}).

Supergranular and granular flows, overshoots of convection cells from below the Sun's surface, drive these photospheric flux features towards downflow regions at the convergence of three or more cells. These photospheric motions result in the following behaviours being displayed by the flux features \citep{cep-Schrijver1997b}: (i) emergence, in which pairs of features, with equal, but opposite-polarity flux, appear; (ii) cancellation, the disappearance (generally through submergence) of equal amounts of flux from a pair of opposite-polarity features; (iii) coalescence, the merging of two like-polarity features creating a larger feature and (iv) fragmentation, the splitting of a large feature into two or more smaller features. Clearly, all these changes in photospheric flux cause changes to the intermingled magnetic loops in the atmosphere above leading to the redistribution of flux between features -- that is these flows and flux changes drive magnetic reconnection \citep{cep-Close2004,cep-Close2005}.

To estimate the time it takes to completely redistribute all the flux
between the features during solar minimum,
\citet{cep-Close2004,cep-Close2005} considered a twelve-hour series of
high-resolution MDI magnetograms in which all photospheric magnetic
features were identified and then tracked in time. Their birth
mechanism (emergence or fragmentation) was noted, as was their death
mechanism (cancellation or coalescence). Potential field
extrapolations were then used to determine the connectivity of the
photospheric flux features. By assuming that the evolution of the
field went through a series of equi-potential states, the observed
connectivity changes were coupled with the birth and death information
of the features to determine the coronal flux recycling/reconnection
time. Remarkably, it was found that during solar minimum the total
flux in the solar corona completely changes all its connections in
just 1.4 hrs \citep{cep-Close2004,cep-Close2005}. A factor of ten
times faster than the time it takes for all the flux in the quiet-Sun
photosphere to be completely replaced
\citep{cep-Schrijver1997b,cep-Hagenaar2003}.

Clearly, reconnection operates on a wide range of scales from kinetic to MHD. The micro-scale physics at the kinetic scales governs the portioning of the released energy into its various new forms and plays a role in determining the rate of reconnection. MHD (the macro-scale physics) determines where the reconnection takes place and, hence where the energy is deposited, and also effects the reconnection rate. In this paper, we focus on macro-scale effects, and investigate the behaviour of three-dimensional (3D) reconnection using MHD numerical experiments.

Two-dimensional (2D) reconnection has been studied in detail and is relatively well understood, especially in the solar and magnetospheric contexts. Over the past decade, our knowledge of 3D reconnection has significantly improved \citep{cep-Lau1990,cep-Priest1995,cep-Demoulin1996b,cep-Priest1996,cep-Birn1998,cep-Longcope2001a,cep-Hesse2001,cep-Pritchett2001,cep-Priest2003,cep-Linton2003,cep-Pontin2006,cep-DeMoortel2006a,cep-DeMoortel2006b,cep-Pontin2007,cep-Haynes2007,cep-Parnell2008a}. It is abundantly clear that the addition of the extra dimension leads to many differences between 2D and 3D reconnection. In Section 2, we first review the key characteristics of both 2D and 3D reconnection. Then, in Section 3, we consider a series of 3D MHD experiments in order to investigate where, how and at what rate reconnection takes place in 3D. The effects of varying resistivity and the resulting energetics of these experiments are discussed in Section 4. Finally, in Section 5, we draw our conclusions.

\section{Characteristics of 2D and 3D reconnection} \label{parnell-sec:characteristics}

\begin{table}
\caption[]{A comparison of the main characteristics of reconnection in 2D and 3D.\label{parnell-table:characteristics}}
\centering
\begin{tabular}{lcl}
\hline
\hspace{1.cm}2D Reconnection &  $\;\;$ &  \hspace{1.cm}3D Reconnection \\
\hline
&&\\[-1.8ex]
1. Must occur at X-type null points &  $\;$ &  1. Can occur at null points or in the \\
 & $\;$ & $\;\;\;$ absence of null points \\
2. Occurs at a single point & $\;$ &  2. Occurs continually and continuously \\
 $\;\;\;$ & $\;$ &   $\;\;\;$ throughout diffusion region volume \\
& $\;$ &   $\;\;\;$ -- not at a single point \\
3. Pairs of field lines break and  &  $\;$ &  3. Pairs of field lines or even pairs of \\
 $\;\;\;$ recombine into two new pairs of  &  $\;$ &   $\;\;\;$ surfaces break, but do not recombine \\

 $\;\;\;$ field lines &  $\;$ &   $\;\;\;$ into two new pairs of field lines or \\
& & $\;\;\;$ surfaces \\
4. Discontinuous field line mapping &  $\;$ &  4. Continuous or discontinuous field line \\
 & & $\;\;\;$ mapping \\ 
5. Stagnation type flow &  $\;$ &  5. Counter-rotating flows \\
\hline
\end{tabular}
\end{table}
A comparison of the main properties of reconnection in 2D and 3D highlight the significant differences that arise due to the addition of the extra dimension (Table~\ref{parnell-table:characteristics}). In 2D, magnetic reconnection can only occur at X-type nulls. Here, pairs of field lines with different connectivities, say $A\rightarrow A'$ and $B\rightarrow B'$ are reconnected at a single point to form a new pair of field lines with connectivities $A\rightarrow B'$ and $B\rightarrow A'$. Hence, flux is transferred from one pair of flux domains into a different pair of flux domains. The fieldline mapping from $A\rightarrow A'$ onto $A\rightarrow B'$ is discontinuous and jumps at the instant of reconnection itself. There is been a considerable body of work on 2D reconnection and a good review of this work can be found in \citet{cep-Priest2000}.

In 3D, magnetic reconnection can occur both at 3D nulls, but more commonly it will occur in a null-less region of magnetic field, for instance, in a hyperbolic flux tube \citep{cep-Galsgaard2003,cep-Linton2003,cep-Pontin2005c,cep-Aulanier2006,cep-DeMoortel2006a,cep-DeMoortel2006b,cep-Wilmot-Smith2007a}, in an elliptic flux tube \citep{cep-Wilmot-Smith2007b} or near a separator \citep{cep-Longcope1996,cep-Galsgaard1997,cep-Galsgaard2000b,cep-Haynes2007,cep-Parnell2008a}. Here, the diffusion region where reconnection occurs is not a single point, but is a finite volume. 

Many, but not all, of the above situations depend on the fact that field lines from flux domains with two different connectivities, say $A\rightarrow A'$ and $B\rightarrow B'$, are reconnected to form field lines in two new flux domains with connectivities $A\rightarrow B'$ and $B\rightarrow A'$, exactly as in the 2D case. In 3D, however, it is generally not possible to identify pairs of field lines (or even pairs of surfaces) that reconnect to form new pairs of field lines (or surfaces). Instead, reconnection will occur continually and continuously throughout the finite diffusion region converting flux from two domains into flux in two other domains. A consequence of having continual and continuous reconnection is that in 3D, the field line mapping can be continuous between pre- and post-reconnected field lines. The theory behind this behaviour is explained in \citet{cep-Hornig2003} and is illustrated very nicely using numerical experiments in \citet{cep-Pontin2005c} and \citet{cep-Aulanier2006}.

Using theoretical arguments, \citet{cep-Hornig2003} determined that counter rotating flows are an essential ingredient of 3D reconnection. \citet{cep-Wilmot-Smith2007a} and \citet{cep-Parnell2009} have both found counter-rotating flows about the reconnection sites in their numerical experiments.

\begin{figure}
\begin{center}
\resizebox{!}{3.9cm}{\includegraphics{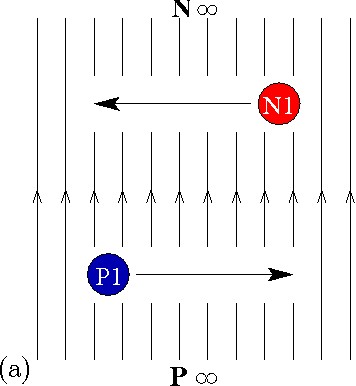}}
\resizebox{!}{3.9cm}{\includegraphics{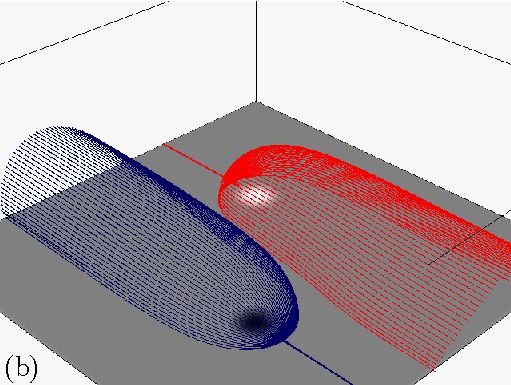}}
\caption[]{(a) Sketch of the experimental setup showing the two opposite-polarity sources, $P1$ \& $N1$ and the overlying field due to $P\infty$ \& $N\infty$. The arrows indicate the direction of advection of $P1$ and $N1$. (b) A three-dimensional view of the initial magnetic field showing fieldlines in the positive (blue) and negative (red) separatrix surfaces.}
\label{parnell-fig:exper}
\end{center}
\end{figure}

\section{3D magnetic interactions}    \label{parnell-sec:3Dmagneticinteractions}

In order, to determine how reconnection occurs in 3D, we investigated a simple interaction which may be considered as a basic building block of the Sun's complex coronal magnetic field. The interaction involves two opposite-polarity magnetic features which are driven past each other in an overlying field \citep{cep-Galsgaard2000,cep-Parnell2004,cep-Galsgaard2005,cep-Haynes2007,cep-Parnell2008a,cep-Haynes2009,cep-Parnell2009}. One would imagine that the resulting magnetic interaction is trivial, after all there are only four sources: $P1$ \& $N1$ -- the two opposite-polarity sources prescribed on the base, and $P\infty$ \& $N\infty$ -- the sources that produce the overlying magnetic field (Fig.~\ref{parnell-fig:exper}a). However, the interaction turns out to be much more complicated than one initially imagines.

\begin{figure*}[t!]
\resizebox{3.85cm}{!}{\includegraphics{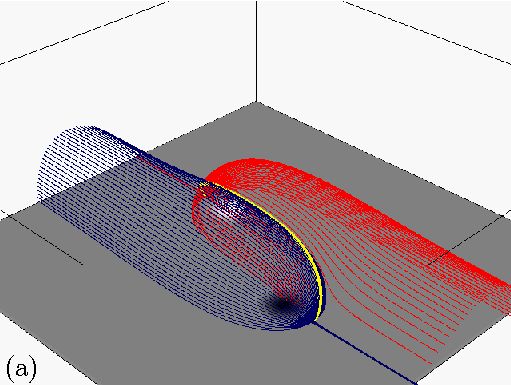}}
\resizebox{3.85cm}{!}{\includegraphics{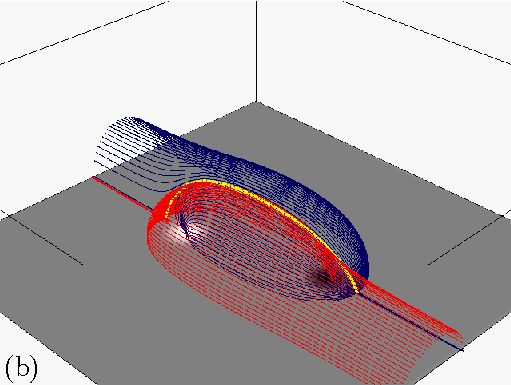}}
\resizebox{3.85cm}{!}{\includegraphics{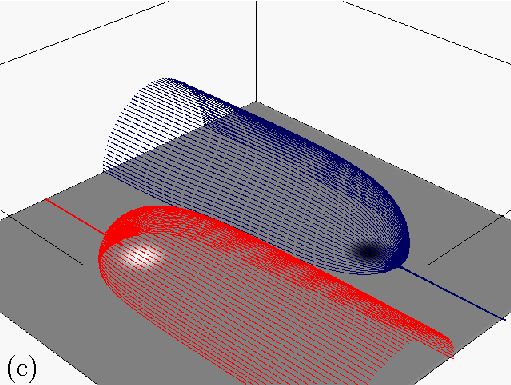}}
\caption[]{Three-dimensional views of the potential magnetic topology evolution during the interaction of two opposite-polarity features in an overlying field: (a) single-separator closing phase; (b) single-separator opening phase; and (c) final phase. Field lines lying in the separatrix surfaces from the positive (blue) and negative (red) nulls are shown. The yellow lines indicate the separators.
}
\label{parnell-fig:pot_skeleton}
\end{figure*}

\subsection{Potential 3D interaction}    \label{parnell-sec:pot3Dinteraction}

For comparative purposes, we first discuss the details of the potential 3D interaction \citep{cep-Haynes2007}. The topology of the initial magnetic field involves just three flux domains: one domain linking $P1$ to $N\infty$, another linking $P\infty$ to $N1$ and the overlying field which links $P\infty$ to $N\infty$, as seen in Fig.~\ref{parnell-fig:exper}b.
In each phase, throughout the evolution, there are two null points: one positive and the other negative. Both of these nulls are situated on the base. From each of these null points a separatrix-surface emanates. In Fig.~\ref{parnell-fig:exper}b, which shows the topology of the {\it initial phase}, the separatrix surfaces are mapped out by the field lines lying in them. The blue(red) field lines show the separatrix surface originating from the positive(negative) null. 

The two opposite-polarity magnetic sources $P1$ and $N1$ are driven in an anti-parallel manner (Fig.~\ref{parnell-fig:exper}a). In this subsection, we assume that they evolve through a series of equi-potential states. This means that the different flux domains interact (reconnect) the moment the separatrix surfaces come into contact. Hence, the first change to a new magnetic topology (new phase) starts as soon as the flux domains from $P1$ and $N1$ come into contact. When this happens a new flux domain and a separator (yellow curve) are created (Fig.~\ref{parnell-fig:pot_skeleton}a). We call this phase the {\it single-separator closing phase}, because the reconnection at this separator transfers flux from the open $P1-N\infty$ and $P\infty-N1$ domains to the newly formed closed, $P1-N1$, domain and the overlying, $P\infty-N\infty$, domain.

When the sources $P1$ and $N1$ reach the point of closest approach all the flux from them has been completely closed and they are fully connected. This state was reached via a global separatrix bifurcation. As they start moving away from each other, the closed flux starts to re-open and a new phase is entered (Fig.~\ref{parnell-fig:pot_skeleton}b). Again, there is still only one separator, but reconnection at this separator now re-opens the flux from the sources (i.e., flux is transferred from the closed, $P1-N1$, and overlying, $P\infty-N\infty$, domains to the two newly formed re-opened, $P1-N\infty$, and, $P\infty-N1$, domains). This is known as the {\it single-separator re-opening phase}.

 Eventually, the two sources $P1$ and $N1$ become completed unconnected from each other, leaving them each just connected to a single source at infinity, and surrounded by overlying field (Fig.~\ref{parnell-fig:pot_skeleton}c). In this phase, the {\it final phase}, there are no separators and there is no reconnection. The field is basically the same as that in the initial phase, but, the two sources ($P1$ \& $N1$) and their associated separatrix surfaces and flux domains have swapped places.

\begin{figure*}[ht]
\resizebox{3.85cm}{!}{\includegraphics{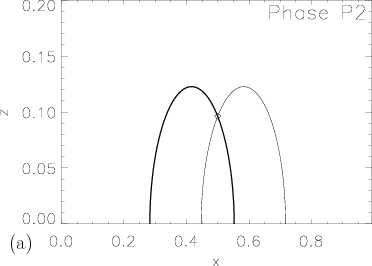}}
\resizebox{3.85cm}{!}{\includegraphics{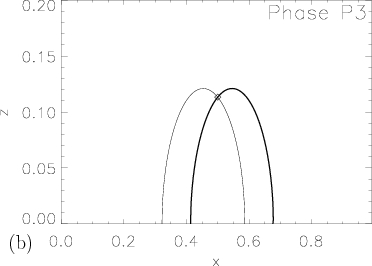}}
\resizebox{3.85cm}{!}{\includegraphics{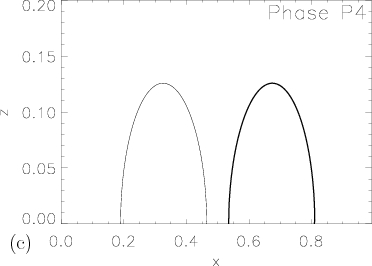}}
\caption[]{Cross-sectional cuts ($y=0.5$ plane) showing the potential magnetic topology evolution during the interaction of two opposite-polarity features in an overlying field. A cut thought the positive separatrix surface is shown by a thick line whilst the cut through the negative one is given by a thin line. Diamonds indicate where the separators intersect the plane. 
}
\label{parnell-fig:pot_skeleton_cut}
\end{figure*}
In order, to visualise the above flux domains, and therefore the magnetic evolution more clearly, we plot 2D cuts taken in the $y=0.5$ planes (Fig.~\ref{parnell-fig:pot_skeleton_cut}). In the three frames of this figure, there are no field lines lying in the plane. Instead, the thick and thin lines show the intersections of the positive and negative separatrix surfaces, respectively, with the $y=0.5$ plane. Where these lines cross there will be a separator threading the plane, shown by a diamond. These frames clearly show the numbers of flux domains and separators during the evolution of the equi-potential field. They are useful as they enable us to easily determine the direction of reconnection at each separator by looking at which domains are growing or shrinking.

Here, the evolution involves $0\rightarrow 1\rightarrow 1\rightarrow 0$ separators, during the four topological phases found. However, such an equi-potential evolution can only arise in a situation where there is perfect, instantaneous reconnection, i.e., in the limit that the magnetic resistivity, $\eta$, tends to $\infty$. This is not very realistic and so, in the subsection below, we consider what happens when we have a more realistic finite resistivity. 

\subsection{3D resistive MHD interaction}    \label{parnell-sec:3DMHDinteraction}
A series of experiments with different constant resistivity (constant-$\eta$) values was considered to see how the evolution of the magnetic topology of the above magnetic interaction varied with varying resistivity \citep{cep-Haynes2009}. A 3D resistive MHD code was used to model these interactions. The initial magnetic setup was exactly that used in the potential case (Fig.~\ref{parnell-fig:exper}b) and the sources were driven in the same manner (Fig.~\ref{parnell-fig:exper}a). Details of the code used and the setup can be found in \citet{cep-Haynes2009}.

\begin{table}
\begin{center}
\caption[]{\label{parnell-table:ceta_evolution}The start times of each of the phases through which the magnetic topology of the various constant resistivity experiments evolve. Each phase is numbered, with the number of separators and numbers of flux domains given in brackets next to the phase number. $S$ is the average maximum Lunquist number of each experiment. The average mean Lunquist number is a factor of 8 smaller than this value. $R_T$ is the number of times that the total flux in a single source reconnects. $\eta_0 = 5\times10^{-4}$.}
\begin{tabular}{c|c|ccccccc|c}
\hline
Res. & $S$ & \multicolumn{7}{c|}{Phases {\it (No. Separators:No. Flux Domains)}} & $R_T$ \\
& & 1{\it (0:3)} & 2{\it (2:5)} & 3{\it (1:4)} & 4{\it (5:8)} & 5{\it (3:6)} & 6{\it (1:4)} & 7{\it (0:3)} &  \\
\hline
&&&&&&&&&\\[-1.8ex]
Pot. & 0 & 0.0 & - & 0.45 & - & - & 4.11 & 7.76 & 2.0 \\ 
$\eta_0$ & $4.8\times 10^3$ & 0.0 & - & 1.50 & - & 6.04 & 7.02 & 10.3 & 2.31 \\ 
$\eta_0/2$ & $9.8\times 10^3$ & 0.0 &- & 1.78 & - & 6.46 & 8.79 & 11.7 & 2.68 \\ 
$\eta_0/4$ & $2.0\times 10^4$ & 0.0 & 1.92 & 2.07 & - & 6.89 & 10.9 & 13.6 & 3.01 \\ 
$\eta_0/8$ & $3.9\times 10^4$ & 0.0 & 2.21 & 2.35 & 7.17 & 7.32 & 14.2 & 16.0 & 3.47 \\
$\eta_0/16$ & $7.9\times 10^4$ & 0.0 & 2.35 & 2.92 & 7.60 & 7.88 & 18.9 & 19.2 & 3.94 \\
\hline
\end{tabular}
\end{center}
\end{table}  
For each constant-$\eta$ experiment, the evolution of the resulting magnetic topology was determined in the same way as above. The different phases found were numbered and described in terms of the numbers of flux domains and separators found, as well as the direction of reconnection at each separator. Table~\ref{parnell-table:ceta_evolution} details the topological evolution found in each constant-$\eta$ experiment. 

As described above, the equi-potential interaction simply evolves though four phases. From Table~\ref{parnell-table:ceta_evolution}, however it is clear, that in the case of finite resistivity, there are always more than four phases. Furthermore, it is also clear that as the resistivity decreases, the number of phases increases, as does the overall duration of the interaction. Below we describe the evolution of the $\eta_0/16$ magnetic topology, since this experiment covers all the phases found.

\begin{figure*}[t!]
\resizebox{3.85cm}{!}{\includegraphics{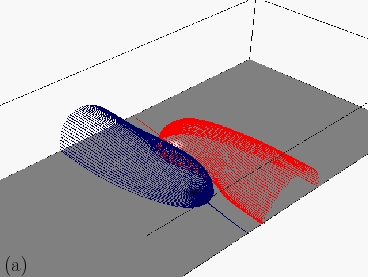}}
\resizebox{3.85cm}{!}{\includegraphics{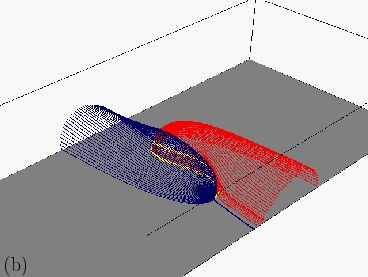}}
\resizebox{3.85cm}{!}{\includegraphics{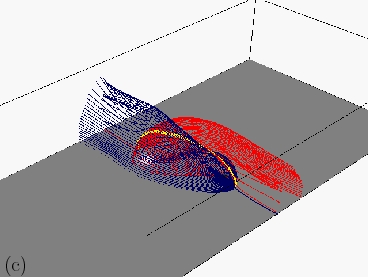}}
\resizebox{3.85cm}{!}{\includegraphics{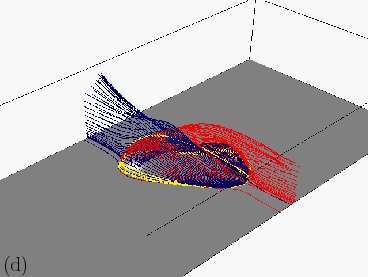}}
\resizebox{3.85cm}{!}{\includegraphics{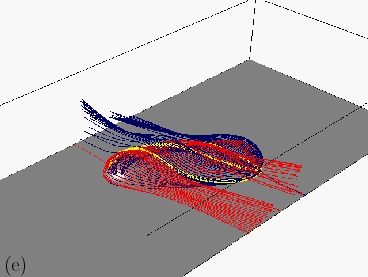}}
\resizebox{3.85cm}{!}{\includegraphics{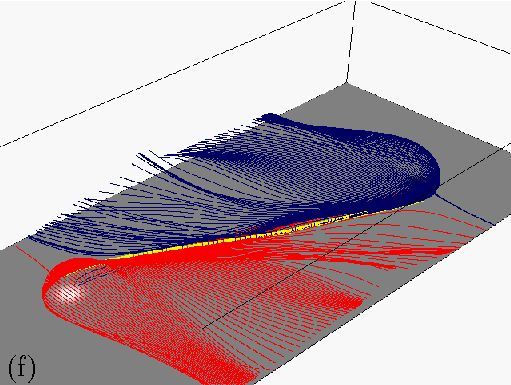}}
\caption[]{Three-dimensional views of the magnetic topology evolution during the $\eta_0/16$ constant-$\eta$ interaction of two opposite-polarity features in an overlying field. Fieldlines in the separatrix surfaces from the positive (blue) and negative (red) are shown. The yellow lines indicate the separators.
}
\label{parnell-fig:skeleton}
\end{figure*}
\begin{figure*}[ht]
\resizebox{3.85cm}{!}{\includegraphics{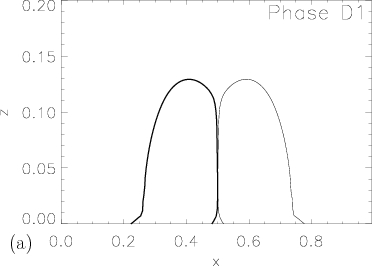}}
\resizebox{3.85cm}{!}{\includegraphics{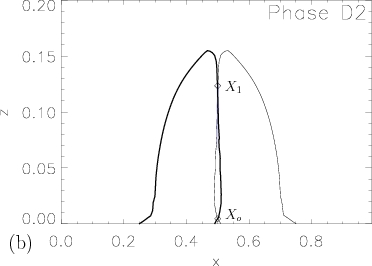}}
\resizebox{3.85cm}{!}{\includegraphics{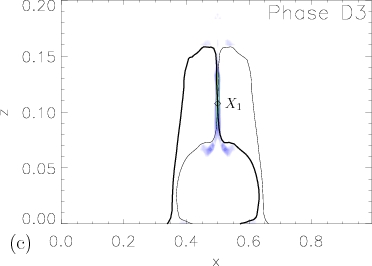}}
\resizebox{3.85cm}{!}{\includegraphics{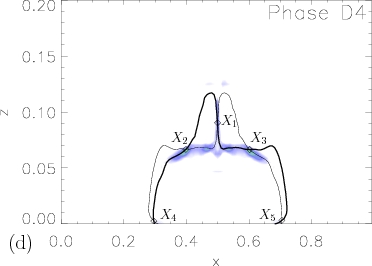}}
\resizebox{3.85cm}{!}{\includegraphics{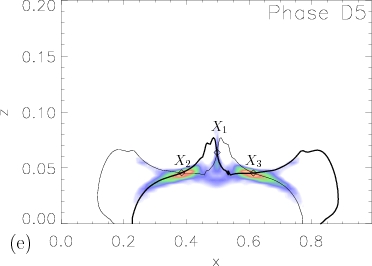}}
\resizebox{3.85cm}{!}{\includegraphics{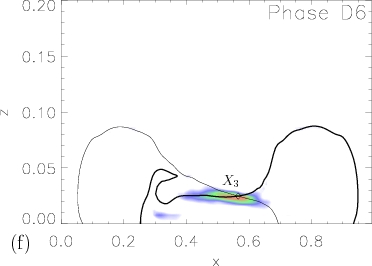}}
\caption[]{The $y=0.5$ cross-sectional cuts through the frames in Fig.~\ref{parnell-fig:skeleton} showing the magnetic topology evolution. Cut though the positive(negative) separatrix surface is shown by a thick(thin) line. Diamonds indicate where the separators intersect the plane. The coloured contours, white -- blue -- green -- red -- yellow, represent increasing current density.}
\label{parnell-fig:skeleton_cut}
\end{figure*}

The magnetic topology of the $\eta_0/16$ experiment evolves through seven different topological phases. Figure~\ref{parnell-fig:skeleton} shows a frame from each of the first six phases. The seventh phase is the same as the final phase seen in the potential evolution and shown in Fig.~\ref{parnell-fig:skeleton}c. For clarity, we also include cross-sections of the magnetic skeleton ($y=0.5$ cuts) for each of these six frames in Fig.~\ref{parnell-fig:skeleton_cut}.  From these two figures, it is clear that the separatrix surfaces intersect each other multiple times giving rise to multiple separators. Also, the filled contours of current in these cross-sections clearly demonstrate that the current sheets in the system are all threaded by a separator. Hence, the number of reconnection sites is governed by the number of separators in the system. 

Figures~\ref{parnell-fig:skeleton}a and \ref{parnell-fig:skeleton_cut}a show the magnetic topology towards the end of the {\it initial phase}, when the sources $P1$ and $N1$ are still unconnected. To enter a new phase reconnection must occur, producing closed flux. Closed flux, connects $P1$ to $N1$ and so must be contained within the two separatrix surfaces, hence these separatrix surfaces must overlap. In the potential situation, the surfaces first overlapped in photosphere (at the base) and so a single intersection (separator) line between the surfaces was found. However, in general, when two surfaces intersect, they will cross in two locations, forming two separators. Figures~\ref{parnell-fig:skeleton}a and \ref{parnell-fig:skeleton_cut}a, show that the two surfaces come closest together above the photosphere (base) and so when they first intersect a new pair of separators ($X_0$ \& $X_1$) is created (via a global double-separator bifurcation (GDSB)) along with two new flux domains: one containing closed flux; the other, trapped below this, containing overlying field (Figs.~\ref{parnell-fig:skeleton}b and \ref{parnell-fig:skeleton_cut}b). This is known as the {\it double-separator hybrid phase}. It is called a hybrid phase, because flux is both closed and re-opened during this phase: the top separator, $X_1$, closes the flux from the sources, whilst reconnection at the lower separator, $X_0$, re-opens this closed flux again, with the reopened flux being transferred into the original open flux domains, $P1-N\infty$ and $P\infty-N1$. The reconnection at the lower separator is very weak and, since the flux in the trapped overlying field domain is small, it soon runs out of flux to reconnect and so disappears, leading to the start of another phase.

Figures~\ref{parnell-fig:skeleton}c and \ref{parnell-fig:skeleton_cut}c show the magnetic topology in this phase, which involves four flux domains and a single separator, $X_1$, at which flux is closed. This is equivalent to the {\it single-separator closing phase} that occurred during the potential evolution. During this phase the closed flux domain increases in size, whilst the two original open flux domains decrease in size. From Fig.~\ref{parnell-fig:skeleton_cut}c, it is clear that the growing closed flux domain pushes the left-hand side of the negative separatrix surface towards the left-hand side of the positive separatrix surface. Similarly, the right-hand side of the positive separatrix surface is pushed towards the right-hand side of the negative separatrix surface. When these surfaces touch a new phase begins.

The new phase starts when flux starts to be re-opened, and this occurs following the occurrence of another two GDSBs, creating four new separators ($X_2$ -- $X_5$) and four new domains. The new separators and domains are created as the inner separatrix surface sides bulge out through the sides of the outer separatrix surfaces. These new separators and flux domains can be clearly seem in Figs.~\ref{parnell-fig:skeleton}d and \ref{parnell-fig:skeleton_cut}d. In total there are eight flux domains and five separators. This phase is called the {\it quintuple-separator hybrid phase}, since flux is both closing and re-opening during this phase. The central separator is separator $X_1$ and reconnection here is still closing flux. Reconnection at separators $X_2$ and $X_3$ (the two upper side separators) is re-opening flux and so filling the two new flux domains below these separators and the original open flux domains above them. At the two lower side separators, $X_4$ and $X_5$, flux is being closed. Below these two separators are two new flux domains, which have been pinched off from the two original open flux domains. Above them are the new re-opened flux domains. It is the flux from these domains that is converted at $X_4$ and $X_5$ into closed flux and overlying flux. These lower side separators do not last long and disappear as soon as the flux in the domains beneath them is used up, which leads to the main reopening flux phase. 

The next phase is called the {\it triple-separator hybrid phase}, and is a phase that occurs in all the constant-$\eta$ experiments (Figs.~\ref{parnell-fig:skeleton}e and \ref{parnell-fig:skeleton_cut}e). 
There is a total of six flux domains and three separators: the central separator ($X_1$) where flux is closed; the side separators ($X_2$ and $X_3$) where flux is re-opened.

The above phase ends, and a new phase starts, when the flux in one of the original open flux domains is used up. This leads to the destruction of separators $X_1$ and $X_2$ via a GDSB, leaving just separator $X_3$, which continues to re-open the remain closed flux (Figs.~\ref{parnell-fig:skeleton}f and \ref{parnell-fig:skeleton_cut}f). This phase is the same as the {\it single-separator re-opening phase} seem in the equi-potential evolution and it ends once all the closed flux has been reopened. The final phase, as has already been mentioned, is the same as that in Figs.~\ref{parnell-fig:pot_skeleton}c and \ref{parnell-fig:pot_skeleton_cut}c and involves no reconnection.

\subsection{Recursive reconnection and reconnection rates} \label{parnell-sec:recursive}
%
\begin{figure}[t!]
\begin{center}
\resizebox{3.0875cm}{!}{\includegraphics{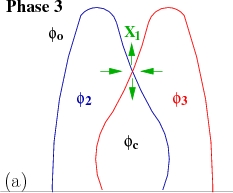}}
\resizebox{3.75cm}{!}{\includegraphics{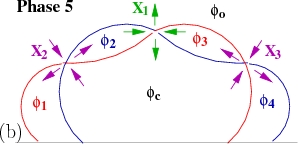}}
\resizebox{3.75cm}{!}{\includegraphics{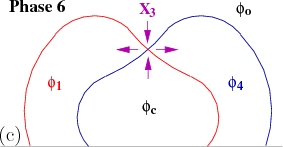}}
\end{center}
\caption[]{Sketch showing the direction of reconnection at (a) the separator, $X_1$ in phase 3, (b) each of the separators, $X_1$-$X_3$ in phase 5 and (c) the separator, $X_3$, in phase 6.}
\label{parnell-fig:recon}
\end{figure}
From Table~\ref{parnell-table:ceta_evolution}, it is clear that there are three main phases involving reconnection in each of the constant-$\eta$ experiments: the {\it single-separator closing phase} (phase 3), the {\it triple-separator hybrid phase} (phase 5) and the {\it single-separator re-opening phase} (phase 6). Figure~\ref{parnell-fig:recon}a shows a sketch of the direction of reconnection at the separator ($X_1$) in phase 3. In this phase, the rate of reconnection across $X_1$ can be simply calculated from the rate of change of flux in anyone of the four flux domains (flux in domains: $\phi_c$ -- closed, $\phi_o$ -- overlying, $\phi_2$ -- original positive open, $\phi_3$ -- original negative open). Hence, the rate of reconnection at $X_1$ during this phase, $\alpha_1$, is given by
$$\alpha_1 = \frac{{\rm d}\phi_c}{{\rm d}t}= -\frac{{\rm d}\phi_{2}}{{\rm d}t}= -\frac{{\rm d}\phi_{3}}{{\rm d}t} = \frac{{\rm d}\phi_o}{{\rm d}t}\;.$$  

Figure~\ref{parnell-fig:recon}b illustrates the direction of reconnection at each of the three separators during phase 5. Here, once again the flux is being closed at the central separator ($X_1$) but at the two outer separators ($X_2$ and $X_3$) it is being re-opened. This overlapping of the two reconnection processes allows flux to both close and then re-open multiple times, i.e., to be {\it recursively reconnected}. There are some interesting consequences from this recursive reconnection, which are discussed below.

Here, the rate of reconnection at the separators $X_2$ and $X_3$  can be simply determined and is equal to
$$\alpha_2 = \frac{{\rm d}\phi_{1}}{{\rm d}t}\;,\;\;\;\;{\rm and}\;\;\;\;\alpha_3 = \frac{{\rm d}\phi_{4}}{{\rm d}t}\;,$$
where $\phi_1$ and $\phi_4$ are the fluxes in the new re-opened negative and positive flux domains, respectively.
The rate of reconnection at $X_1$ is slightly harder to determine since every flux domain surrounding this separator is losing, as well as gaining flux. The rate of reconnection, $\alpha_1$, during this phase equals
$$\alpha_1 = \frac{{\rm d}\phi_{1}}{{\rm d}t}-\frac{{\rm d}\phi_{2}}{{\rm d}t}= \frac{{\rm d}\phi_{4}}{{\rm d}t}-\frac{{\rm d}\phi_{3}}{{\rm d}t}\;.$$

Figure~\ref{parnell-fig:recon}c illustrates the direction of reconnection at the separator $X_3$ during phase 6, the {\it single separator re-opening phase}. 
Here, the rate of reconnection, $\alpha_3$ at separator $X_3$ is simply equal to
$$\alpha_3 = -\frac{{\rm d}\phi_{c}}{{\rm d}t} = \frac{{\rm d}\phi_{4}}{{\rm d}t}= \frac{{\rm d}\phi_{1}}{{\rm d}t} = -\frac{{\rm d}\phi_{o}}{{\rm d}t}\;.$$

\begin{figure}[t!]
\begin{center}
\resizebox{11cm}{!}{\includegraphics*{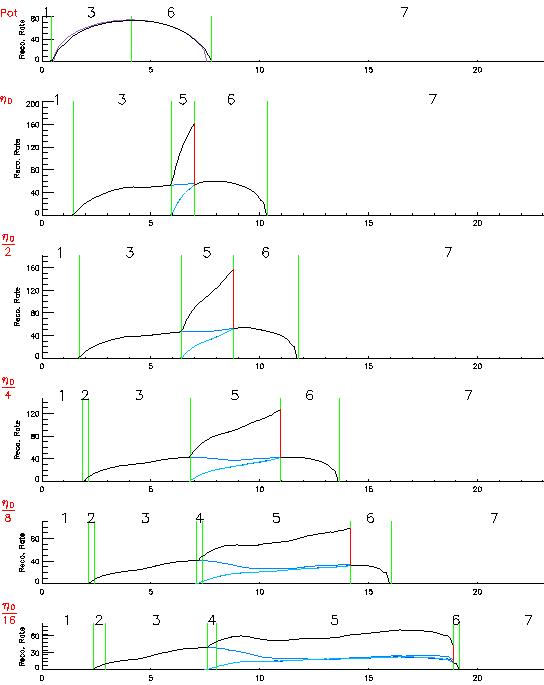}}
\end{center}
\caption[]{Plots of the global reconnection rates against time for the equi-potential evolution, as well as each of the constant-$\eta$ experiments. The black lines show the total reconnection rate, the red-lines show the discontinuous jumps that occur upon the destruction of pairs of separators, and the blue lines show the reconnection rates at individual separators.}
\label{parnell-fig:grr}
\end{figure}
For each experiment, it is possible to calculate the global rate of reconnection in the experiment, $\alpha$,
$$\alpha=\sum_{i=0}^{5}\alpha_i\;,$$
where $\alpha_i=0$ when the separator $X_i$ does not exist. Plots of the global reconnection rate, $\alpha$, against time for each experiment are shown in Fig.~\ref{parnell-fig:grr}, with the start and end of each phase labelled. From these graphs, we note the following points: (i) as the value of $\eta$ decreases, the instantaneous reconnection rate falls, with the peak rate in the $\eta_0$ experiment some 2.4 times greater than the peak rate in the $\eta_0/16$ experiment, and (ii) as $\eta$ decreases, the overall duration of the interaction increases. 

By integrating under the black curves, we can calculate the total reconnection in each experiment,
$$R_T = \int_0^\infty \alpha \;{\rm d}t\;.$$
These numbers are given in the right-hand column of Table~\ref{parnell-table:ceta_evolution} and it is clear that $R_T$ increases as $\eta$ decreases ($R_T(\eta_0/16)/R_T({\rm pot}) = 1.97$). 

Where does this ``extra flux'' come from that is reconnected? In the potential case, there is no overlap between the closing and reopening phases and the total flux from one source is simply reconnected twice through the following evolution:
$${\rm open}\;\;\rightarrow\;\;{\rm closed}\;\;\rightarrow\;\;{\rm reopened}\;.$$
In the case of finite $\eta$, there is an overlap between the closing and re-opening reconnection processes, which permits recursive reconnection, hence the flux now evolves in the following manner:
$${\rm open}\;\;\rightarrow\;\;{\rm closed}\;\;\rightleftarrows\;\;{\rm reopened}\;.$$
Thus, there is no actual extra flux, instead re-opened flux is closed and re-opened again multiple times during phase 5.

From Fig.~\ref{parnell-fig:grr} we see that phase 3, which only closes the flux, lasts for about the same duration in each of the constant-$\eta$ experiments, although the flux closed during this time decreases as $\eta$ decreases. Phase 5, however, increases in length considerably as $\eta$ decreases, permitting more and more recursive reconnection. In contrast, phase 6, the simple reopening phase, decreases in duration, such that in the $\eta_0/16$ case it lasts less than half an Alfv{\'e}n crossing time.  

During phase 5, there is a rapid increase in the global reconnection rate (Fig.~\ref{parnell-fig:grr}). The rapidity of this increase decreases as $\eta$ decreases. The upper blue line plotted during this phase shows the rate of reconnection at separator $X_1$, whilst the lower blue line shows the rate of reconnection at both separators $X_2$ and $X_3$. When these blue lines meet, all three separators are reconnecting at exactly the same rates. Since, the system is symmetric, the rates of reconnection at all three separators are always essentially the same at the end of phase 5, at which point two of the separators coincide and destroy each other. Hence, the discontinuous jump at the end phase 5 (red line) always involves the global rate of reconnection dropping by a factor of three.
 
\section{Energetics}                   \label{parnell-sec:energetics}

We have seen that the instantaneous global rate of reconnection drops as $\eta$ decreases, but the total flux reconnected increases. So what are the implications for the rate of heating and the total amount of free magnetic energy of the interaction?

\begin{figure}[t!]
\begin{center}
\resizebox{3.75cm}{!}{\includegraphics{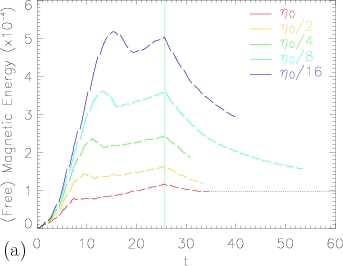}}
\resizebox{3.75cm}{!}{\includegraphics{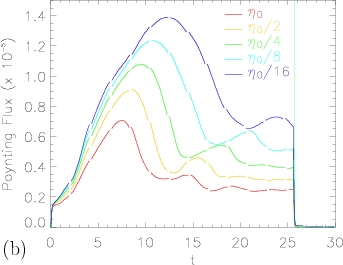}}
\resizebox{3.75cm}{!}{\includegraphics{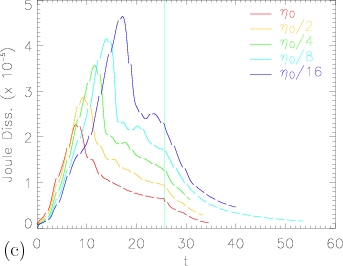}}
\end{center}
\caption[]{(a) Free magnetic energy, (b) Poynting flux and (c) ohmic dissipation against time for the constant-$\eta$ experiments. The green vertical line indicates the end of the driving of the sources.}
\label{parnell-fig:nrg}
\end{figure}
As $\eta$ decreases, the total amount of free magnetic energy in the system increases (Fig.~\ref{parnell-fig:nrg}a). Although, not shown in this graph, we find that all experiments eventually relax down to the same magnetic energy. The extra energy in the small constant-$\eta$ experiments is injected through the base as Poynting flux (Fig.~\ref{parnell-fig:nrg}b). This is because, as $\eta$ decreases, the onset of reconnection is delayed (Fig.~\ref{parnell-fig:grr}) and so the magnetic field from the sources becomes more horizontal. Therefore, the normal component of the Poynting flux, $(E_xB_y-E_yB_x)/\mu$, increases. 

Figure~\ref{parnell-fig:nrg}c, indicates that as $\eta$ decreases, not only does the total energy released from the system increase, but the peak rate of joule dissipation also increases. This means that, although the instantaneous rate of reconnection in the system decreases, the rate of heating increases. This is because the heating from the reconnection (ohmic heating) is proportional $\eta j^2$. So the decrease in $\eta$ is offset by a far greater increase in the current density. The currents are built up due to the delay in onset of reconnection caused by the small value of $\eta$.       

\section{Conclusions}                   \label{parnell-sec:conclusion}

Research over the last decade has revealed that the properties of 3D magnetic reconnection are quite different to those of 2D reconnection. In particular, 3D reconnection can occur in both null and non-null regions. Furthermore, it does not occur at a single point, but occurs continuously and continually throughout the diffusion volume. 

Through the detailed analysis of a simple 3D magnetic interaction, forming a basic building block of the Sun's magnetic atmosphere, we have investigated some of the interesting consequences of 3D reconnection. In particular, we have demonstrated that the reconnection within a simple magnetic flux system is not trivial, but is instead highly complex. The key to revealing the details of the interaction was understanding the evolution of the magnetic topology. i.e.,  determining the evolution of the magnetic skeleton. This allowed us to determine: 
\begin{itemize} \vspace*{-1ex}
\item the location of the reconnection/energy release sites,
\item the direction of the reconnection processes that occurred at these sites,
\item the rates of reconnection at these sites and also the global reconnection rate.
\end{itemize}

An unexpected discovery from the magnetic skeleton analysis was the process of recursive reconnection, where the same flux can be reconnected multiple times. 
It is interesting to consider what structures permit the recursive reconnection to take place. Recursive reconnection is possible because several separators link the same two null points: the nulls are said to be {\it multiply connected} \citep{cep-Parnell2007}. Multiply-connected nulls go hand-in-hand with multiply-connected source pairs. They can occur in all types of magnetic fields: potential \citep{cep-Parnell2007}, force-free, and non force-free, as seen here. 

The process of recursive reconnection leads to the following important consequences:
\begin{itemize} \vspace{-1ex}
\item {\it Greater spread of energy}: Recursive reconnection involves many separators at which reconnection is occurring, and hence many energy release sites. This leads to a better/wider distribution of energy which is important, for instance, for heating the corona. The more energy release sites there are the easier it is to maintain a background temperature of over a million degrees across almost the whole surface.
\item {\it Repeated heating of the plasma}: Since flux is repeatedly reconnected, it is possible for certain parts of the plasma to have injections of energy in rapid succession over a short period of time enabling it to potentially get significantly hotter than other regions of plasma around it. Furthermore, this repeated reconnection of certain parts of flux may explain why, in some circumstances, you can observe solar flare loops before they reconnect when you would normally expect them to be cool and so invisible in X-rays. 
\item {\it Longer period of heating}: Since flux is repeatedly closed and re-opened many times, the entire heating process lasts longer than if it simply closed and re-opened just once. This again is useful for coronal heating and also solar flares.
\end{itemize}

Finally, it is interesting to observe that the rate of reconnection is not proportional to the rate of ohmic heating. In fact, here, as $\eta$ decreases, so does the rate of reconnection, but the rate and peak value of ohmic heating increases. Further work needs to be undertaken to see how these two rates are related in general.  


\begin{small}


\bibliographystyle{rr-assp}       

\end{small}

\end{document}